\documentclass
[nofootinbib,preprint,12pt,prd,groupedaddress,showpacs,eqsecnum,titlepage,showkeys]{revtex4}%
\usepackage{amsfonts}
\usepackage{amsmath}
\usepackage{amssymb}
\usepackage{graphicx}
\usepackage{latexsym}
\usepackage{color}%
\begin{document}
\title{Dilatonic effects on a falling\ test mass in scalar-tensor theory}
\author{J.R. Morris}
\affiliation{Physics Dept., Indiana University Northwest, 3400 Broadway, Gary, Indiana
46408, USA}
\email{jmorris@iun.edu}

\begin{abstract}
Effects of a 4d dilaton field on a falling test mass are examined from the
Einstein frame perspective of scalar-tensor theory. Results are obtained for
the centripetal acceleration of particles in circular orbits, and the radial
acceleration for particles with pure radial motion. These results are applied
to the specific case of nonrelativistic motion in the weak field approximation
of Brans-Dicke theory, employing the exact Xanthopoulos-Zannias solutions. For
a given parameter range, the results obtained from Brans-Dicke theory are
qualitatively dramatically different from those of general relativity.
Comments are made concerning a comparison with the general relativistic
results in the limit of an infinite Brans-Dicke parameter.

\end{abstract}

\maketitle

\section{\bigskip Introduction}

Scalar-tensor theories form a class of candidates for a modified description
of gravity, and some type of modified gravity at large distances could give
rise to observable deviations from general relativity. Brans-Dicke
theory\cite{Brans-Dicke} is a prototypical scalar-tensor theory where, in a
Jordan frame representation, a massless scalar field couples nonminimally to
the Ricci curvature scalar. However, more general scalar-tensor theories allow
different couplings of the scalar \textquotedblleft dilaton\textquotedblright%
\ field to the curvature, as well as accommodating nonzero scalar field
potentials. Four dimensional scalar-tensor theories arise from a variety of
theoretical approaches aimed at achieving unification and/or explaining
certain types of observations. Such approaches include Kaluza-Klein type
models, string theory, and brane-world models involving extra space
dimensions, and result in effective four dimensional models of gravity with a
nonminimally coupled scalar field\cite{F-M}. Therefore, a study of the effects
presented by a general form of scalar-tensor theory will include the effects
that emerge from a variety of higher dimensional models, as well as four
dimensional scalar-tensor theories that may not require extra dimensions.

\bigskip

A fairly general form of a scalar-tensor theory is considered here, and we
concentrate on the Einstein frame representation of the theory where the
dilatonic effects and the metric tensor field effects can be distinguished
more easily. We then proceed to find expressions for the motion of a test
particle moving in a static, spherically symmetric background. Expressions are
obtained for (1) the angular speed of a test mass in circular motion, and (2)
the radial acceleration of a particle undergoing pure radial motion.
Simplification results when we consider nonrelativistic motion. As an example,
we apply these expressions to the exact analytical vacuum solutions of
Brans-Dicke theory\cite{Brans-Dicke}, i.e., the Xanthopoulos-Zannias
solutions\cite{X-Z}, which solve the Einstein frame field equations. The
differences between the Brans-Dicke results and the general relativity (GR)
results are seen, and for a given parameter range, are dramatically different
in a qualitative sense. Comments are also offered to illustrate in a concrete
way, that, as pointed out by Faraoni\cite{Faraoni1},\cite{Faraoni2}, when the
matter stress-energy vanishes, GR is not generically recovered from the
Brans-Dicke theory in the limit of an infinite Brans-Dicke parameter.

\section{Conformal frames}

Consider a Jordan frame representation of a scalar-tensor theory of the form%
\begin{equation}
S=\int d^{4}x\sqrt{\tilde{g}}\left\{  \frac{F(\tilde{\phi})}{2\kappa^{2}%
}\tilde{R}[\tilde{g}_{\mu\nu}]+\frac{1}{2}\tilde{g}^{\mu\nu}\partial_{\mu
}\tilde{\phi}\partial_{\nu}\tilde{\phi}-V(\tilde{\phi})\right\}  +S_{m}%
[\tilde{g}_{\mu\nu}] \label{1}%
\end{equation}

where $\kappa^{2}=16\pi G$, $\tilde{g}=|\det\tilde{g}_{\mu\nu}|$, the scalar
field $\tilde{\phi}$ is identified as a 4d dilaton with a potential
$V(\tilde{\phi})$, and a metric signature $(+,-,-,-)$ is used. The Jordan
frame metric and line element are given by $d\tilde{s}^{2}=\tilde{g}_{\mu\nu
}dx^{\mu}dx^{\nu}$. The matter action $S_{m}[\tilde{g}_{\mu\nu}]$ is
constructed from the metric $\tilde{g}_{\mu\nu}$ and matter terms. For
instance, a classical particle action can be written as%
\begin{equation}
S_{m,cl}=-\sum_{A}\int m_{0,A}d\tilde{s}_{A}=-\sum_{A}\int m_{0,A}\left[
\tilde{g}_{\mu\nu}(x_{A})dx_{A}^{\mu}dx_{A}^{\nu}\right]  ^{1/2} \label{1a}%
\end{equation}

where $m_{0,A}$ is the mass of particle $A$ in the Jordan frame, assumed to be
a constant. A field theoretic matter action is%
\begin{equation}
S_{m}=\int d^{4}x\sqrt{\tilde{g}}\mathcal{\tilde{L}}_{m}(\tilde{g}_{\mu\nu
},\psi) \label{1b}%
\end{equation}

where $\psi$ labels matter fields. A classical matter Lagrangian density can
be defined by\cite{Dam-Poly},\cite{JM09}%
\begin{equation}
\sqrt{\tilde{g}}\mathcal{\tilde{L}}_{cl}=-\sum_{A}\int m_{0,A}\left[
\tilde{g}_{\mu\nu}(x_{A})dx_{A}^{\mu}dx_{A}^{\nu}\right]  ^{1/2}\delta
^{(4)}(x-x_{A}) \label{1c}%
\end{equation}

The associated stress-energy tensors for field theoretic or classical actions%
\begin{equation}
\mathcal{\tilde{T}}^{\mu\nu}=\frac{2}{\sqrt{\tilde{g}}}\frac{\partial
(\sqrt{\tilde{g}}\mathcal{\tilde{L}}_{m})}{\partial\tilde{g}_{\mu\nu}%
},\ \ \ \ \ \mathcal{\tilde{T}}_{cl}^{\mu\nu}=-\frac{2}{\sqrt{\tilde{g}}}%
\frac{\partial(\sqrt{\tilde{g}}\mathcal{\tilde{L}}_{cl})}{\partial\tilde
{g}_{\mu\nu}}\label{1d}%
\end{equation}

then give $\mathcal{\tilde{T}}_{00}>0$ in both cases.

\bigskip

The Einstein frame representation of this theory is obtained with a rescaling
of the metric and scalar field\cite{Dicke62},\cite{Kaiser},\cite{Faraoni} :%
\begin{equation}
\tilde{g}_{\mu\nu}\rightarrow g_{\mu\nu}=\Omega^{2}\tilde{g}_{\mu\nu
},\ \ \ \ \Omega=\sqrt{F(\tilde{\phi})},\ \ \ \ \tilde{\phi}\rightarrow
\phi(\tilde{\phi}),\ \ \frac{d\phi}{d\tilde{\phi}}=\frac{1}{F}\left\{
F+\frac{3}{16\pi G}\left[  F^{\prime}(\tilde{\phi})\right]  ^{2}\right\}
^{1/2}\ \label{2}%
\end{equation}

where $F^{\prime}(\tilde{\phi})=dF/d\tilde{\phi}$, giving an Einstein frame
representation%
\begin{equation}
S=\int d^{4}x\sqrt{g}\left\{  \frac{1}{2\kappa^{2}}R[g_{\mu\nu}]+\frac{1}%
{2}g^{\mu\nu}\partial_{\mu}\phi\partial_{\nu}\phi-U(\phi)\right\}
+S_{m}(\Omega^{-2}g_{\mu\nu}) \label{3}%
\end{equation}

The potential $U(\phi)$ depends upon the functions $F(\tilde{\phi})$ and
$V(\tilde{\phi}(\phi))$,%
\[
U(\phi)=\frac{V}{\Omega^{4}}=\frac{V[\tilde{\phi}(\phi)]}{F^{2}[\tilde{\phi
}(\phi)]}%
\]
(See, for example, \cite{Kaiser}.) The Einstein frame line element is
$ds^{2}=g_{\mu\nu}dx^{\mu}dx^{\nu}=\Omega^{2}d\tilde{s}^{2}=F(\tilde{\phi
})d\tilde{s}^{2}$. In the Einstein frame a particle has a mass $m$, which is
generally position dependent due to its dependence on the scalar field
$\tilde{\phi}$. Consider, for example, a classical matter action of the form
in (\ref{1a}),%
\begin{equation}
-S_{m}=\int m_{0}d\tilde{s}=\int m_{0}(\Omega^{-1}ds)=\int m_{0}%
F^{-1/2}ds=\int mds \label{4}%
\end{equation}

so that the Einstein frame mass $m$ is related to the Jordan frame mass
$m_{0}$ by\cite{Dicke62},\cite{Kaiser}%
\begin{equation}
m=\Omega^{-1}m_{0}=F^{-\frac{1}{2}}(\tilde{\phi})m_{0} \label{5}%
\end{equation}

Therefore, a particle having a constant mass $m_{0}$ in the Jordan frame will
have a mass $m=F^{-1/2}m_{0}$ in the Einstein frame. Since the fields
$\tilde{\phi}$ and $\phi$ generally depend on spacetime position, then the
Einstein frame mass $m=m(x^{\mu})$ in general. The matter Lagrangian density
in the Einstein frame, $\mathcal{L}_{m}$, is related to that in the Jordan
frame, $\mathcal{\tilde{L}}_{m}$, by\cite{Dicke62},\cite{Kaiser}%
\begin{equation}
\mathcal{L}_{m}=\Omega^{-4}\mathcal{\tilde{L}}_{m}(\tilde{g}_{\mu\nu}%
)=F^{-2}\mathcal{\tilde{L}}_{m}(\tilde{g}_{\mu\nu}) \label{6}%
\end{equation}

\bigskip

A particular example is that of Brans-Dicke (BD) theory\cite{Brans-Dicke},
with a Jordan frame action ($G=1$)%
\begin{equation}
S=\frac{1}{16\pi}\int d^{4}x\sqrt{\tilde{g}}\left\{  \tilde{\phi}\tilde
{R}+\frac{\omega_{BD}}{\tilde{\phi}}\tilde{g}^{\mu\nu}\partial_{\mu}%
\tilde{\phi}\partial_{\nu}\tilde{\phi}\right\}  +S_{m}(\tilde{g}_{\mu\nu})
\label{7}%
\end{equation}
\bigskip A conformal transformation to the Einstein frame is given
by\cite{Cai-Myung}%
\begin{equation}
g_{\mu\nu}=\tilde{\phi}\tilde{g}_{\mu\nu},\ \ g^{\mu\nu}=\tilde{\phi}%
^{-1}\tilde{g}^{\mu\nu},\ \ \ \sqrt{g}=\tilde{\phi}^{2}\sqrt{\tilde{g}%
},\ \ \ \phi=\sqrt{2a}\ln\tilde{\phi},\ \ \ a=\omega_{BD}+\frac{3}{2}
\label{8}%
\end{equation}

and the action in the Einstein frame then takes the form%
\begin{equation}
S=\frac{1}{16\pi}\int d^{4}x\sqrt{g}\left\{  R+\frac{1}{2}g^{\mu\nu}%
\partial_{\mu}\phi\partial_{\nu}\phi\right\}  +S_{m}(\tilde{\phi}^{-1}%
g_{\mu\nu}) \label{9}%
\end{equation}

where $R$ is built from $g_{\mu\nu}$ and Einstein gravity is coupled to a
massless Einstein frame scalar dilaton field $\phi$. Using $g_{\mu\nu}%
=\Omega^{2}\tilde{g}_{\mu\nu}$ as in (\ref{2}), we identify $\Omega
=\tilde{\phi}^{1/2}$ and from (\ref{5}) we have%
\begin{equation}
m=\Omega^{-1}m_{0}=\tilde{\phi}^{-1/2}m_{0} \label{10}%
\end{equation}

(The kinetic term in (\ref{7}) is in noncanonical form, but a rescaling of the
scalar field\cite{Kaiser} $\tilde{\phi}\rightarrow\bar{\phi}^{2}/(8\omega
_{BD})$ would put the kinetic term into a canonical form as in (\ref{1}), with
$F(\bar{\phi})\propto\bar{\phi}^{2}/(8\omega_{BD})$.) Terms in the matter
Lagrangian $\mathcal{L}_{m}=\Omega^{-4}\mathcal{\tilde{L}}_{m}(\tilde{g}%
_{\mu\nu})=\tilde{\phi}^{-2}\mathcal{\tilde{L}}_{m}(\tilde{g}_{\mu\nu})$ pick
up an anomalous coupling to the dilaton $\tilde{\phi}$.

\bigskip

\bigskip

A classical test particle of mass $m$ moving in a gravitational field
described by $ds^{2}=g_{\mu\nu}dx^{\mu}dx^{\nu}$ has an action like that in
(\ref{4}),%
\begin{equation}
S=-\int m\left[  g_{\mu\nu}u^{\mu}u^{\nu}\right]  ^{1/2}ds \label{11}%
\end{equation}

where $u^{\alpha}=dx^{\alpha}/ds$ is subject to the \textquotedblleft on
shell\textquotedblright\ constraint $u_{\alpha}u^{\alpha}=1$. The
\textquotedblleft geodesic\textquotedblright\ equation of the motion (in an
otherwise matter-free region) obtained from (\ref{11}) can be written in the
form\cite{Dicke62}%
\begin{equation}
\frac{d}{ds}\left(  mg_{\mu\nu}u^{\nu}\right)  -\frac{1}{2}m(\partial_{\mu
}g_{\alpha\beta})u^{\alpha}u^{\beta}-\partial_{\mu}m=0 \label{12}%
\end{equation}

or in the form
\begin{equation}
\frac{du^{\nu}}{ds}=-\Gamma_{\alpha\beta}^{\nu}u^{\alpha}u^{\beta}%
+\partial_{\mu}(\ln m)(g^{\mu\nu}-u^{\mu}u^{\nu}) \label{13}%
\end{equation}

The first term on the right hand side of (\ref{13}) is recognized as the
gravitational acceleration due to the metric field $g_{\mu\nu}$, while the
second term on the right hand side represents the dilatonic acceleration due
to the scalar field, and therefore a deviation from pure, unforced, geodesic
motion. Since $m(x^{\mu})\propto\Omega^{-1}(x^{\mu})=F^{-1/2}(x^{\mu})$, the
motion of a particle in the Einstein frame of a scalar-tensor theory where
$\partial_{\mu}m\neq0$, will differ from that described by general relativity
(GR) where $m=$ const. This reflects the fact that the Jordan frame metric
$\tilde{g}_{\mu\nu}$ for a scalar-tensor theory will generally be different
from the metric of GR. Since the acceleration of a test mass in the Einstein
frame depends upon the tensor field $g_{\mu\nu}$ as well as the dilatonic
acceleration due to the scalar field $\phi$, it is not enough to consider the
asymptotic form of the metric alone, e.g., $g_{00}-1$, for the case of an
asymptotically flat spacetime.

\section{Motion in a static, spherically symmetric background}

We now focus upon the motion of a classical test particle of mass $m(r)$
moving under the influence of a metric field $g_{\mu\nu}$ in the Einstein
frame of a scalar-tensor theory that can be written in the form
of\ eq.(\ref{1}). We assume that $g_{\mu\nu}$ and $m$ are static and
spherically symmetric functions, independent of $t$ and azimuth angle
$\varphi$, with $g_{\mu\nu}$ being diagonal, and consider motion in the
equatorial plane, $\theta=\pi/2$. The special cases of circular motion and
pure radial motion will be considered by using (\ref{12}) or (\ref{13}), along
with the constraint $u_{\alpha}u^{\alpha}=1$. Different coordinate systems can
be used (Schwarzschild-like or isotropic), but we take the metric to have a
general form%
\begin{equation}
ds^{2}=e^{f(r)}dt^{2}-e^{-h(r)}dr^{2}-\rho(r)r^{2}d\Omega^{2} \label{14}%
\end{equation}

where $\rho(r)=e^{-h(r)}$ for isotropic coordinates, and $d\Omega^{2}%
=d\theta^{2}+\sin^{2}\theta d\varphi^{2}$.

\bigskip

First, we point out that if the motion is initially within the equatorial
plane $\theta=\pi/2$, then it remains in this plane, so that $u^{\theta
}=d\theta/ds=0$. This is seen from the $\theta$ component of (\ref{12}), which
reduces to $\frac{d(mu_{\theta})}{ds}-\frac{m}{2}\rho(r)r^{2}\sin\theta
\cos\theta(u^{\varphi})^{2}=0$, so that if $\theta=\pi/2$ and $u_{\theta}=0$
initially, then $d\left(  mu_{\theta}\right)  /ds=0$ initially (no $\theta$
component of acceleration), so that motion remains in the $\theta=\pi/2$ plane.

\bigskip

For the $t$ component equation, $\partial_{0}g_{\alpha\beta}=0$ and
$\partial_{0}m=0$, so $\frac{d}{ds}(m\,u_{0})=0$, which gives%
\begin{equation}
p_{0}=mu_{0}=E;\ \ \ \ \ u^{0}=\frac{E}{mg_{00}} \label{15}%
\end{equation}

where the energy $E$ is a constant parametrizing the particular orbit. For
example, a given circular orbit has a fixed value of $E$, but this value will
generally depend upon the orbital radius $r$, so that $E=E(r)$ is a constant
on the orbit. For pure radial motion, $E$ might characterize the asymptotic
energy of the test mass, and different values of $E$ characterize different
radial orbits (e.g., different turning points).

\bigskip

Similarly, for the $\varphi$ equation, $\frac{d}{ds}(mu_{\varphi})=0$, with%
\begin{equation}
p_{\varphi}=mu_{\varphi}=mg_{\varphi\varphi}u^{\varphi}=-L;\ \ \ \ u_{\varphi
}=\frac{-L}{m},\ \ \ \ u^{\varphi}=\frac{-L}{mg_{\varphi\varphi}} \label{16}%
\end{equation}

where the angular momentum $L$ is a constant of motion characterizing the
orbit, and $g_{\varphi\varphi}$ is evaluated at $\theta=\pi/2$ for circular
orbits. For pure radial motion, $L=0$, but for a circular orbit $L$ depends
upon the orbital radius, as with Newtonian gravity.

\bigskip

The radial equation reduces to%
\begin{equation}
\frac{d}{ds}(mu_{r})-\frac{1}{2}m\left[  (\partial_{r}g_{00})(u^{0}%
)^{2}+(\partial_{r}g_{rr})(u^{r})^{2}+(\partial_{r}g_{\varphi\varphi
})(u^{\varphi})^{2}\right]  -\partial_{r}m=0 \label{17}%
\end{equation}

and the constraint equation is $u_{\alpha}u^{\alpha}=g^{00}(u_{0}%
)^{2}+g^{\varphi\varphi}(u_{\varphi})^{2}+g_{rr}(u^{r})^{2}=1$. Using
(\ref{15}) and (\ref{16}), this constraint gives%
\begin{equation}
(u^{r})^{2}=\frac{1}{|g_{rr}|}\left[  \frac{E^{2}}{g_{00}m^{2}}-\frac{L^{2}%
}{|g_{\varphi\varphi}|m^{2}}-1\right]  \label{18}%
\end{equation}

The kinematically allowed regions where the test mass can propagate are
defined by $(u^{r})^{2}\geq0$, with radial turning points given by
$(u^{r})^{2}=0$.

\bigskip

We could also define an effective potential $\mathcal{V}$ for the radial
motion by (see, for example,\cite{Ohanian-Ruffini},\cite{PQR}) $(u^{r}%
)^{2}+\mathcal{V}^{2}=E^{2}/m^{2}$, where%
\begin{equation}
\mathcal{V}^{2}=\frac{1}{|g_{rr}|}\left(  1+\frac{L^{2}}{|g_{\varphi\varphi
}|m^{2}}\right)  +\frac{E^{2}}{m^{2}}\left(  1-\frac{1}{g_{00}|g_{rr}%
|}\right)  \label{19}%
\end{equation}

with radial turning points determined by $E/m=\mathcal{V}$.

\subsection{Circular motion}

For circular motion $u^{r}=0$ and $L$ and $E$ are constants of the particular
orbit. We write%
\begin{equation}
u^{\varphi}=\frac{-L}{mg_{\varphi\varphi}}=\frac{d\varphi}{ds}=u^{0}%
\frac{d\varphi}{dt}=u^{0}\omega=\frac{E}{mg_{00}}\omega\label{20}%
\end{equation}

where the angular speed $\omega=d\varphi/dt$, and we have used (\ref{15}) and
(\ref{16}). Therefore the angular speed $\omega$ can be written as%
\begin{equation}
\omega=\frac{L}{E}\frac{g_{00}}{|g_{\varphi\varphi}|} \label{21}%
\end{equation}

with $E$ and $L$ related by (\ref{18}) with $u^{r}$ set to zero. For example,
consider the case of nonrelativistic circular motion of a test particle with
constant mass $m$ due to Newtonian gravity in a Minkowski spacetime, where
$g_{00}\rightarrow1$ and $|g_{\varphi\varphi}|\rightarrow r^{2}$ in the
equatorial plane, and $E\rightarrow m$. We then have from (\ref{21})
$\omega=L/mr^{2}$, the ordinary Newtonian relation.

\bigskip

Now the relation between $E$ and $L$, given by (\ref{18}) with $u^{r}$ set to
zero, is equivalent to the constraint equation $p_{\mu}p^{\mu}=m^{2}$ with
$p^{r}=0$, and leads to%
\begin{equation}
\frac{E}{m}=\sqrt{g_{00}}\left[  1+\frac{L^{2}}{|g_{\varphi\varphi}|m^{2}%
}\right]  ^{1/2} \label{22}%
\end{equation}

The second term within brackets on the right is recognized as an orbital
kinetic energy (per unit mass) term. Eq. (\ref{22}) allows $E$ to be
eliminated from the expression in (\ref{21}), leaving $\omega$ to be
determined by $L$, along with the metric $g_{\mu\nu}$ and mass function
$m(r)$. The radial equation (\ref{17}),%
\begin{equation}
(\partial_{r}g_{00})\left(  \frac{E}{mg_{00}}\right)  ^{2}+(\partial
_{r}g_{\varphi\varphi})\left(  \frac{L}{mg_{\varphi\varphi}}\right)
^{2}=-2\frac{\partial_{r}m}{m} \label{23}%
\end{equation}

then allows a determination of $L$ in terms of the $g_{\mu\nu}$ and $m$. Thus
$\omega$, evaluated on the orbit with radius $r$, will ultimately depend not
only upon the metric $g_{\mu\nu}$, but also upon the mass function $m(r)$ and
its rate of change $\partial_{r}m$, evaluated on the orbit with radius $r$.

\subsubsection{\textbf{Nonrelativistic limit: }$|\vec{p}|/m\ll1$\textbf{:}}

The above procedure and eq.(\ref{21}) simplifies in the nonrelativistic limt
of low velocities, $v\ll1$, or where the particle kinetic energy is much
smaller than its mass energy, $|p_{i}p^{i}|\ll m^{2}$. Then $p_{\mu}p^{\mu
}=g^{00}E^{2}+p_{i}p^{i}=m^{2}\approx g^{00}E^{2}$, or%
\begin{equation}
E\approx m\sqrt{g_{00}} \label{24}%
\end{equation}

(This is also obtained from (\ref{22}) when we drop the orbital kinetic energy term.)

\bigskip

The radial equation (\ref{23}), with the use of (\ref{24}) and some
rearrangement, yields%
\begin{equation}
\frac{L}{m}\approx|g_{\varphi\varphi}|\left\{  \frac{1}{\partial
_{r}|g_{\varphi\varphi}|}\partial_{r}\left[  \ln(m^{2}g_{00})\right]
\right\}  ^{1/2} \label{25}%
\end{equation}

in the nonrelativistic limit. Using (\ref{24}) and (\ref{25}), the angular
speed in (\ref{21}) is given in the nonrelativistic limit by%
\begin{equation}
\omega^{2}\approx\frac{\partial_{r}(m^{2}g_{00})}{m^{2}\partial_{r}%
|g_{\varphi\varphi}|} \label{26}%
\end{equation}

To test this, we use the Schwarzschild solution in Schwarzschild coordinates,
with $m=$ const, $g_{00}=1-2GM/r$, and $|g_{\varphi\varphi}|=r^{2}$ in the
equatorial plane. Eq.(\ref{26}) then gives the Newtonian limit, $\omega
^{2}=GM/r^{3}$ and a centripetal acceleration $|a_{c}|=\omega^{2}r=GM/r^{2}$,
i.e., the ordinary Newtonian gravitational field produced by a static,
spherically symmetric body of mass $M$. However, it is possible that a
scalar-tensor theory with nonconstant mass and a different metric field could
yield a dramatically different result. An example is provided later for
Brans-Dicke theory.

\subsection{Radial Motion}

We take $\theta=\pi/2$, $\varphi=const$, so that $u^{\theta}=u^{\varphi}=0$
for pure radial motion, and the constraint equation becomes $g^{00}(u_{0}%
)^{2}-|g_{rr}|(u^{r})^{2}=1$, with $u_{0}=E/m$. We also have the radial
component of equation (\ref{13}), which becomes%
\begin{equation}
\frac{du^{r}}{ds}=-\left[  \Gamma_{00}^{r}(u^{0})^{2}+\Gamma_{rr}^{r}%
(u^{r})^{2}\right]  +\frac{\partial_{r}m}{m}\left[  g^{rr}-(u^{r})^{2}\right]
\label{27}%
\end{equation}

with $\Gamma_{00}^{r}=-\frac{1}{2}g^{rr}\partial_{r}g_{00}$ and $\Gamma
_{rr}^{r}=\frac{1}{2}g^{rr}\partial_{r}g_{rr}$. The constraint equation%
\begin{equation}
(u^{r})^{2}=\frac{1}{|g_{rr}|}\left[  \frac{E^{2}}{g_{00}m^{2}}-1\right]
\label{28}%
\end{equation}

and $u^{0}=g^{00}E/m$ can be used in (\ref{27}), allowing the proper radial
acceleration to be given by%
\begin{equation}
g_{rr}\frac{du^{r}}{ds}=\frac{1}{2}(\partial_{r}g_{00})\left[  \frac{E^{2}%
}{g_{00}^{2}m^{2}}\right]  -\frac{1}{2}g^{rr}(\partial_{r}g_{rr})\left[
1-\frac{E^{2}}{g_{00}m^{2}}\right]  +\frac{\partial_{r}m}{m}\left[
\frac{E^{2}}{g_{00}m^{2}}\right]  \label{29}%
\end{equation}

\subsubsection{Nonrelativistic limit: $|\vec{p}|/m\ll1$:}

In the nonrelativistic limit $|p_{r}p^{r}|/m^{2}\ll1$, or $g_{rr}(u^{r}%
)^{2}\ll1$, then (\ref{28}) implies that $\frac{E^{2}}{g_{00}m^{2}}\approx1$,
i.e., the condition given by (\ref{24}), and we also have $ds\approx
\sqrt{g_{00}}dt$. The geodesic equation then takes the simplified form%
\begin{equation}
\frac{d^{2}r}{dt^{2}}\approx\left(  \frac{g_{00}}{g_{rr}}\right)  \left[
\frac{1}{2}(\partial_{r}g_{00})+\frac{\partial_{r}m}{m}\right]  \label{30}%
\end{equation}

in describing the radial acceleration of a test mass $m$ in the
nonrelativistic limit, where for a scalar-tensor theory $m=m(r)$ in the
Einstein frame. As an example, we again apply this to the Schwarzschild case,
where $m$ is constant and $g_{00}=-g_{rr}^{-1}=(1-2GM/r)$, to get a radial
acceleration $a_{r}=a_{c}=-GM/r^{2}$, the usual Newtonian limit. However, we
will also consider an example from Brans-Dicke theory.

\section{Application to Brans-Dicke theory}

We now apply the results of (\ref{26}) and (\ref{30}) to the case of a static,
spherically symmetric background of Brans-Dicke (BD) theory. The Jordan frame
representation of the BD theory is given by (\ref{7}). Exact static,
spherically symmetric vacuum solutions in the Jordan frame were provided by
Brans\cite{Brans}. The conformal transformations described by (\ref{8}) allows
the theory to be rewritten in the Einstein frame representation, given by
(\ref{9}). The BD vacuum solutions in the Einstein frame, as well as the
higher dimensional generalizations, have been provided by Xanthopoulos and
Zannias\cite{X-Z}. Cai and Myung\cite{Cai-Myung} have also studied these
solutions, explicitly relating the Jordan frame solutions and the Einstein
frame solutions through the transformations of (\ref{8}). We apply these
solutions to describe the region exterior to some neutral, nonrotating
astrophysical object of BD theory, and look at the asymptotic limit $r\gg
r_{0}$. (There is a naked singularity at $r=r_{0}$, except in the case of the
Schwarzschild limit, where the solution coincides with the Schwarzschild
solution\cite{X-Z},\cite{Cai-Myung}.) However, the solution inside the
astrophysical object will not be a vacuum solution, so that we do not
generally expect a physical singularity to exist. For an astrophysical object
like a star or planet, we expect that $r/r_{0}\gg1$ for all regions outside
the surface.

\bigskip

The static neutral solutions, with isotropic coordinates, are presented here
for the special 4d case:%
\begin{equation}
ds^{2}=e^{f}dt^{2}-e^{-h}(dr^{2}+r^{2}d\Omega^{2})\smallskip\label{31}%
\end{equation}

\begin{equation}
e^{f}=g_{00}=\xi^{2\gamma}\smallskip;\ \ \ \ \ \ \xi=\left(  \dfrac{r-r_{0}%
}{r+r_{0}}\right)  \label{32}%
\end{equation}

\begin{equation}
e^{-h}=|g_{rr}|=\left(  1-\dfrac{r_{0}^{2}}{r^{2}}\right)  ^{2}\xi^{-2\gamma
}=e^{-f}\left(  1-\dfrac{r_{0}^{2}}{r^{2}}\right)  ^{2} \label{33}%
\end{equation}

\begin{subequations}
\label{34}%
\begin{align}
\phi &  =\pm\tilde{\gamma}\ln\xi=\sqrt{2a}\ln\tilde{\phi}\smallskip
;\ \ \ \ \ \tilde{\gamma}=[4(1-\gamma^{2})]^{1/2}\label{34a}\\
\tilde{\phi}  &  =\xi^{\Gamma};\ \ \ \ \Gamma=\pm\frac{\tilde{\gamma}}%
{\sqrt{2a}}=\pm\left[  \frac{2}{a}(1-\gamma^{2})\right]  ^{1/2}\smallskip
=\pm|\Gamma| \label{34b}%
\end{align}

where $r_{0}$ and $\gamma$ are integration constants ($r_{0}>0$), and we have
defined%
\end{subequations}
\begin{equation}
\xi=\left(  \frac{r-r_{0}}{r+r_{0}}\right)  \leq1,\ \ \ \ \ \tilde{\gamma
}=[4(1-\gamma^{2})]^{1/2},\ \ \ \ \ \ \Gamma=\pm\frac{\tilde{\gamma}}%
{\sqrt{2a}}=\pm\left[  \frac{2}{a}(1-\gamma^{2})\right]  ^{1/2} \label{35}%
\end{equation}
These are the Einstein frame fields and solutions, with $0\leq\gamma\leq1$ for
the description of physical (nonegative ADM mass) solutions. There is a naked
singularity at $r=r_{0}$ where $R=g^{\mu\nu}R_{\mu\nu}\rightarrow\infty$
unless $\gamma=1$ and $\phi=0$ (the Schwarzschild solution).

\bigskip

\textit{Note}: In the set of solutions presented in ref.\cite{X-Z}, only the
solution with the $+$ sign in (\ref{34a}), i.e., $\phi=+\tilde{\gamma}\ln\xi$,
is presented. However, the second solution $\phi=-\tilde{\gamma}\ln\xi$ is
seen to exist due to the invariance of the action and equations of motion
(EoM) under the transformations $g_{\mu\nu}\rightarrow g_{\mu\nu}$,
$\phi\rightarrow-\phi$. Thus if $\phi$ is a solution to the EoM, then so is
$-\phi$ (see, for example, refs.\cite{Cai-Myung} and\cite{W-R}). Therefore
$\phi$ can be positive or negative, and the Brans-Dicke scalar $\tilde{\phi
}=\xi^{\Gamma}=\xi^{\pm|\Gamma|}$ can be either a decreasing or an increasing
function of $r$ and $\xi$. The Einstein frame mass $m$ of a test particle is
given by (\ref{10}) and (\ref{34b}),
\begin{equation}
m=m_{0}\tilde{\phi}^{-1/2}=m_{0}\xi^{-\Gamma/2} \label{36}%
\end{equation}

where $m_{0}$ is the constant Jordan frame mass.

\bigskip

We now want to consider the asymptotic forms of these solutions for which
$r_{0}/r\ll1$. In this case we have the following approximations to
$O(r_{0}/r)$.%
\begin{equation}%
\begin{array}
[c]{cc}%
\xi\approx1-2\frac{r_{0}}{r},\ \ \ \ \ g_{00}\approx1-4\gamma\frac{r_{0}}%
{r},\ \ \ \ |g_{rr}|\approx1/g_{00},\medskip & \\
|g_{\varphi\varphi}|_{\theta=\pi/2}\approx r^{2}\left(  1+4\gamma\frac{r_{0}%
}{r}\right)  ,\ \ \ \partial_{r}|g_{\varphi\varphi}|_{\theta=\pi/2}%
\approx2r\left(  1+2\gamma\frac{r_{0}}{r}\right)  ,\medskip & \\
\frac{m^{2}}{m_{0}^{2}}\approx\left(  1+2\Gamma\frac{r_{0}}{r}\right)
,\ \ \ \frac{m^{2}}{m_{0}^{2}}g_{00}\approx1-2(2\gamma-\Gamma)\frac{r_{0}}%
{r}, &
\end{array}
\label{37}%
\end{equation}

For the case of nonrelativistic particle motion, applying (\ref{37}) to
(\ref{26}) for the case of circular motion yields the result%
\begin{equation}
\omega^{2}\approx\left(  2\gamma-\Gamma\right)  \frac{r_{0}}{r^{3}%
},\ \ \ \ \ \ a_{c}=-\omega^{2}r\approx-(2\gamma-\Gamma)\frac{r_{0}}{r^{2}}
\label{38}%
\end{equation}

The Schwarzschild case is obtained for $\gamma=1$, $\Gamma=0$, and the
identification $r_{0}=GM/2$, where $M$ is the mass of the gravitating
object\cite{X-Z},\cite{Ohanian-Ruffini}. The Schwarzschild limit therefore
gives $\omega^{2}\rightarrow GM/r^{3}$ and $a_{c}\rightarrow-GM/r^{2}$, i.e.,
the Newtonian limit of the gravitational field far from the Schwarzschild
radius. (The Schwarzschild radial coordinate $R$ is related to the isotropic
coordinate $r$ by\cite{X-Z},\cite{Ohanian-Ruffini} $R=r\left(  1+r_{0}%
/r\right)  ^{2}$, with $R\rightarrow r$ asymptotically.) Similarly, applying
(\ref{37}) to (\ref{30}) for the case of radial motion, we have%
\begin{equation}
\frac{d^{2}r}{dt^{2}}\approx-(2\gamma-\Gamma)\frac{r_{0}}{r^{2}} \label{39}%
\end{equation}

We therefore have the same gravitational acceleration, $a_{c}=a_{r}$, for
circular or radial motion, with the expected Newtonian limit for the
Schwarzschild case.

\bigskip

A qualitative distinction between GR and BD in the weak field limit is seen
for the case when $\Gamma>2\gamma$, in which case the radial acceleration
$a_{r}\approx(\Gamma-2\gamma)r_{0}/r^{2}$ given by (\ref{39}), becomes
\textit{positive} rather than \textit{negative}, indicating a
\textit{repulsion} rather than an \textit{attraction}. Similarly, (\ref{38})
implies that $\omega^{2}<0$ in this case, i.e., circular orbits do not exist,
implying a \textit{repulsion}, rather than \textit{attraction}. This is seen
as an example where there is a dilatonic repulsion that dominates the metric
field attraction, since, from (\ref{37}), the metric field produces a
$g_{00}-1<0$, and $g_{00}(\partial_{r}g_{00})>0$, but $m(r)$ is a decreasing
function with $\partial_{r}m<0$. Specifically, from (\ref{30}) and (\ref{37}),%
\begin{equation}
\frac{d^{2}r}{dt^{2}}\approx-\left(  \frac{g_{00}}{|g_{rr}|}\right)  \left[
\frac{1}{2}(\partial_{r}g_{00})+\frac{\partial_{r}m}{m}\right]  \approx
(g_{00})^{2}\left[  -\left(  2\gamma\frac{r_{0}}{r^{2}}\right)  +\frac
{|\partial_{r}m|}{m}\right]  ,\ \ \ (\Gamma>2\gamma) \label{40}%
\end{equation}

showing that the metric field (due to the first term on the right) produces a
negative acceleration, but the dilatonic acceleration due to the mass (due to
the second term on the right) produces a positive radial acceleration, which
overwhelms the negative metric contribution. (An interesting occurrence of
repulsive gravity in GR has also been reported\cite{PQR}.)

\bigskip

For the case $\Gamma=2\gamma$, then $\omega^{2}\rightarrow0$ and
$a_{r}\rightarrow0$, so that a test mass at rest remains at rest. However, for
the parameter range $\Gamma<2\gamma$, which is satisfied for $\Gamma>0$ with
$|\Gamma|<2\gamma$, and for all $\Gamma<0$, the radial acceleration is
negative, with an overall attraction.

\bigskip

The solar system constraint on the massless Brans-Dicke theory requires
$\omega_{BD}>40,000$ \cite{Bertotti,Periv}. However,
Perivolaropoulos\cite{Periv} has recently re-examined this constraint for the
case where the Brans-Dicke scalar has an arbitrary nonzero mass $m_{BD}$. The
conclusion reached is that for a mass $m_{BD}\gtrsim200\times10^{-27}$ GeV,
all values of $\omega_{BD}\geq-3/2$ are allowed by solar system observations.
(The extent of the deformation of the Xanthopoulos-Zannias solutions due to a
tiny scalar mass, however, is not apparent.)

\bigskip

It can also be pointed out that in the limit of $\omega_{BD}\rightarrow\infty
$, in which case $a\rightarrow\infty$, $\Gamma\rightarrow0$, and the scalar
field is removed from the theory ($\phi\rightarrow0$ with $\tilde{\phi
}\rightarrow1$), the radial acceleration obtained from the BD theory becomes
$a_{r}\approx-2\gamma r_{0}/r^{2}$, which is a factor of $\gamma$ times the
Schwarzschild value obtained from GR. The reason for this can be seen from the
Xanthopoulos-Zannias solutions (\ref{31}) - (\ref{34}), noting that the metric
$g_{\mu\nu}$ in this case does not collapse to the Schwarzschild metric for
$\gamma\neq1$ \cite{Faraoni2}. This illustrates in a concrete way the point
made by Faraoni\cite{Faraoni1},\cite{Faraoni2} that BD theory does not always
reduce to GR in the $\omega_{BD}\rightarrow\infty$ limit when the matter
stress-energy vanishes, with $\mathcal{T}^{\mu\nu}=0$.

\section{Summary}

A fairly general form of scalar-tensor theory has been considered, with a
focus on the Einstein frame representation of the theory, where scalar field
dilatonic effects and metric tensor field effects become distinguishable.
Expressions for the motion of a test particle moving in a static, spherically
symmetric background are found (1) for the case of circular motion, and (2)
for the case of pure radial motion. Simplified expressions are obtained for
nonrelativistic particle motion. As an example, these expressions have been
applied to the exact analytical vacuum solutions to Brans-Dicke theory, by
using the Xanthopoulos-Zannias solutions for the field equations in the
Einstein frame. The differences between the Brans-Dicke results and the
general relativity results are seen. For a given parameter range, namely, for
$\Gamma>2\gamma$, these are dramatically different qualitatively, as the
dilatonic repulsion of a test mass is greater than the gravitational
attraction due to the tensor field. Furthermore, it is illustrated in a
concrete way that, as pointed out previously by Faraoni\cite{Faraoni1}%
,\cite{Faraoni2}, when the matter stress-energy vanishes, $\mathcal{T}^{\mu
\nu}=0$, GR is not automatically recovered from the Brans-Dicke theory in the
limit of an infinite Brans-Dicke parameter, $\omega_{BD}\rightarrow\infty$.
The Xanthopoulos-Zannias solutions in this limit do not coincide with the
Schwarzschild solution, unless the Xanthopoulos-Zannias parameter is unity,
$\gamma=1$.

\end{document}